\documentclass[a4paper,11pt]{article}
\pdfoutput=1

\usepackage{jcappub}

\usepackage[english]{babel}
\usepackage{booktabs}

\title{\bf Measurement of the cosmic ray spectrum 
above $\mathbf{4{\times}10^{18}}$\,eV using inclined
events detected with the Pierre Auger Observatory}

\collaboration{The Pierre Auger Collaboration}

\author[41]{A.~Aab} 
\author[65]{P.~Abreu} 
\author[52]{M.~Aglietta} 
\author[82]{E.J.~Ahn} 
\author[28]{I.~Al Samarai} 
\author[16]{I.F.M.~Albuquerque} 
\author[1]{I.~Allekotte} 
\author[87]{P.~Allison} 
\author[11,\: 8]{A.~Almela} 
\author[58]{J.~Alvarez Castillo} 
% ...

\affiliation[1]{ Centro At\'{o}mico Bariloche and Instituto Balseiro (CNEA-UNCuyo-CONICET), San 
Carlos de Bariloche, 
Argentina }
% ...
\affiliation[8]{ Instituto de Tecnolog\'{\i}as en Detecci\'{o}n y Astropart\'{\i}culas (CNEA, CONICET, UNSAM), 
Buenos Aires, 
Argentina }
% ...
\affiliation[11]{ Universidad Tecnol\'{o}gica Nacional - Facultad Regional Buenos Aires, Buenos Aires,
Argentina }
% ...
\affiliation[16]{ Universidade de S\~{a}o Paulo, Instituto de F\'{\i}sica, S\~{a}o Paulo, SP, 
Brazil }
% ...
\affiliation[28]{ Institut de Physique Nucl\'{e}aire d'Orsay (IPNO), Universit\'{e} Paris 11, CNRS-IN2P3, 
Orsay, 
France }
% ...
\affiliation[41]{ Universit\"{a}t Siegen, Siegen, 
Germany }
% ...
\affiliation[52]{ Osservatorio Astrofisico di Torino  (INAF), Universit\`{a} di Torino and Sezione INFN, 
Torino, 
Italy }
% ...
\affiliation[58]{ Universidad Nacional Aut\'{o}noma de M\'{e}xico, M\'{e}xico, D.F., 
M\'{e}xico }
% ...
\affiliation[65]{ Laborat\'{o}rio de Instrumenta\c{c}\~{a}o e F\'{\i}sica Experimental de Part\'{\i}culas - LIP and  
Instituto Superior T\'{e}cnico - IST, Universidade de Lisboa - UL, 
Portugal }
% ...
\affiliation[82]{ Fermilab, Batavia, IL, 
USA }
% ...
\affiliation[87]{ Ohio State University, Columbus, OH, 
USA }
% ...

\emailAdd{auger\_spokespersons@fnal.gov}

\abstract{A measurement of the cosmic-ray spectrum for energies exceeding
  $4{\times}10^{18}$\,eV is presented, which is based on the analysis of
  showers with zenith angles greater than $60^{\circ}$ detected with the
  Pierre Auger Observatory between 1 January 2004 and 31 December 2013. The
  measured spectrum confirms a flux suppression at the highest energies. Above
  $5.3{\times}10^{18}$\,eV, the ``ankle'', the flux can be described by a
  power law $E^{-\gamma}$ with index $\gamma=2.70 \pm 0.02 \,\text{(stat)} \pm
  0.1\,\text{(sys)}$ followed by a smooth suppression region. For the energy
  ($E_\text{s}$) at which the spectral flux has fallen to one-half of its
  extrapolated value in the absence of suppression, we find
  $E_\text{s}=(5.12\pm0.25\,\text{(stat)}^{+1.0}_{-1.2}\,\text{(sys)}){\times}10^{19}$\,eV.}

\dedicated{doi:10.1088/1475-7516/2015/08/049}

\begin{document}
\maketitle
\flushbottom

\section{Introduction}
\label{sec:intro}
\normalsize

The Pierre Auger Observatory is to date the largest detector of air showers
induced by the ultra-high energy cosmic rays (UHECRs). It is a hybrid detector
combining an array of surface detectors (SD), that measures the particle
densities in the shower at the ground, and a fluorescence detector (FD), that
captures the ultraviolet light emitted by nitrogen as showers develop in the
atmosphere. The SD~\cite{SD,Aab:2015zoa}, spread over an area of 3000\,km$^2$,
is composed of a baseline array, comprising more than 1600 water-Cherenkov
detectors placed in a hexagonal grid with nearest neighbours separated by
1500\,m. The FD~\cite{FD} comprises 24 fluorescence telescopes distributed in
4 perimeter buildings which view the atmosphere over the array. The site is
located near Malarg\"ue, Province of Mendoza, Argentina, at an altitude of
about 1400\,m above sea level and at an average geographic latitude of
$35.2^\circ$\,S. The redundancy provided by the two detection techniques has
proved to be extremely valuable for a wide range of applications and has
improved the performance of the Observatory beyond expectations.

The data gathered at the Pierre Auger Observatory with zenith angles less than
$60^\circ$ have provided a measurement of the UHECR
spectrum~\cite{SpectrumPRL} with unprecedented statistics. The technique
developed exploits the large aperture of the SD, operating continuously, as
well as the calorimetric measurement of the energy deposit obtained with the
FD which is, by contrast, rather limited in duty cycle to clear nights without
moonlight ($13\%$). A parameter quantifying the shower size is obtained from
the SD. This parameter is then calibrated using the energy inferred from the
calorimetric FD measurement for a sub-sample of the events (hybrid events)
which are detected and reconstructed simultaneously with both
techniques~\cite{SpectrumPRL}. This allows measurements of the energy spectrum
with an energy estimation weakly reliant upon shower simulations. The spectrum
obtained is consistent with the Greisen and Zatsepin-Kuz'min (GZK)
suppression~\cite{GZK1,GZK2}. The observation of this strong flux suppression
was first reported by HiRes~\cite{Hires}. In addition, the spectrum has been
independently measured using hybrid events which are detected with the
fluorescence technique and at least one particle
detector~\cite{HybridSpectrum}. The latter measurement has also been combined
with the SD spectrum to obtain a spectrum extending to lower
energies~\cite{HybridSpectrum,SchulzICRC2013}, which has confirmed the
flattening of the spectral slope at about $5{\times}10^{18}$\,eV, often
referred to as the ``ankle''. Similarly the Telescope Array (TA)
experiment~\cite{TASD,TAFD}, having an array of scintillators
spread over an area of 700\,km$^2$, together with fluorescence telescopes, has
provided an independent measurement of the spectrum using showers with zenith
angles less than $45^\circ$ in the Northern Hemisphere, which clearly displays
the GZK steepening~\cite{TAspectrum}. Although the two spectra are in
agreement between the ankle and $5{\times}10^{19}$\,eV, there are
discrepancies in the suppression region which are being scrutinized by a joint
working group of both Collaborations.

The choice of deep-water Cherenkov detectors for the Pierre Auger Observatory
was made to give substantial solid angle coverage for zenith angles exceeding
$60^\circ$ thus enabling detection of showers over a much broader angular and
declination range than is possible in arrays using thinner detectors. The
showers with large zenith angles traverse much larger atmospheric depths than
than showers nearer the vertical and, as a result, are characterised by the
dominance of secondary energetic muons at the ground, since the
electromagnetic component of the shower is largely absorbed before reachig
ground. Inclined showers have been observed since the 1960s and have stirred
much interest~\cite{Hillas}, but the first reliable methods for event
reconstruction are relatively recent~\cite{Ave,InclinedRec}. These events
provide an independent measurement of the cosmic-ray spectrum.  In addition,
since inclined showers are mainly composed by muons, their analysis has been
shown to be particularly useful in determining the muon content of showers for
comparison with predictions from shower models~\cite{MuonNumberMeasurement}.
Inclined showers are also used to extend the accessible fraction of the sky
covered in analyses of the distribution of the arrival directions of cosmic
rays detected at Auger~\cite{PierreAuger:2014yba,ThePierreAuger:2014nja} and
constitute the main background for searches for ultra-high energy neutrinos
using air showers~\cite{Capelle}.

In this article we present the measurement of the cosmic-ray spectrum derived from
events with zenith angles between $60^\circ$ and $80^\circ$ detected with the
Pierre Auger Observatory in the time period from 1 January 2004 to 31 December
2013.

\section{Estimation of the shower energy}
\label{sec:reconstruction}

Since inclined showers are dominated by muons that are strongly affected by
the geomagnetic field, they require a specific reconstruction procedure that
differs from that used for events with zenith angles less than $60^\circ$. The
method to estimate the energy of an inclined event recorded by the SD array,
described in full detail in~\cite{InclinedRec}, is summarised in the
following.

The arrival direction $(\theta,\phi)$ of the shower is obtained by fitting the
start times of the signals recorded at the triggered stations to a plane front
corrected for small time delays associated with the muon propagation as modelled
in~\cite{LCB}. Once the shower geometry is established, the intensities of the
signals are used to quantify the shower size and the impact point of the
shower axis on the ground. The method is based on the observation that the
shape of the muon distribution is approximately universal for a given shower
direction and that only the overall normalisation of the muon distribution
depends on the shower energy and primary
mass~\cite{InclinedRec,Universality}. It has also been shown that the lateral
shape of the muon density profile is consistently reproduced by different hadronic
models and software packages used for air-shower simulations. These universal
characteristics allow one to model the muon number density as a function of
the position at the ground (muon distribution) as:
\begin{equation}
\rho_{\mu}(\vec{r}) = N_{19}\;\rho_{\mu,19}(\vec{r};\theta,\phi).
\label{MDensity}
\end{equation}
Here $N_{19}$ is a measure of the shower size and is the relative
normalisation of a particular event with respect to a reference muon
distribution, $\rho_{\mu,19}(\vec{r};\theta,\phi)$~\cite{model,billoir}, and
$\rho_{\mu,19}$ is conventionally chosen to be the average muon density for primary protons
with an energy of $10^{19}$\,eV simulated using
\textsc{QGSJetII-03}~\cite{QGSJETII-03} as the hadronic interaction model.
$N_{19}$ is thus expected to be correlated with the shower energy, and
independent of the shower arrival direction. Therefore, $N_{19}$ can be used
as the energy estimator of inclined events recorded at the SD.

$N_{19}$ is estimated by fitting the measured signals at the SD stations to
the expected muon distribution for a given arrival direction, using a
maximum-likelihood method based on a probabilistic model of the detector
response to muon hits obtained from \textsc{Geant4}~\cite{GEANT} simulations
with the
\mbox{$\overline{\textrm{Off}}$\hspace{.05em}\protect\raisebox{.4ex}{$\protect\underline{\textrm{line}}$}}
framework~\cite{Offline} of the Pierre Auger Observatory.  A residual
electromagnetic component of the signal stemming from the muons themselves,
mainly from decay in flight, is taken into account based on simulations
(typically amounting to 20\% of the muon signal)~\cite{Ines}. To ensure a good
reconstruction, only events well-contained in the SD array are selected. This
fiducial trigger requires that the reconstructed core of the shower falls
within an active cell, defined as the elemental hexagon cell\footnote{The
  elemental hexagon cell is the region with vertices at the barycentre of
  each of the six triangles around the central station, with area given by
  $D^2 \sqrt{3}/2$ where $D=1.5$\,km is the array spacing~\cite{Piera}.}
surrounding an operational station when the six neighbouring stations in the
regular hexagonal pattern are also operational, at the time of the event.

Although the actual value for $N_{19}$ depends upon the particular choice of
composition and hadronic model made for the reference distribution, the
absolute energy calibration is inferred from a sub-sample of hybrid events
measured simultaneously with the FD and SD that are used to calibrate the
shower size $N_{19}$ with the calorimetric shower energy measured with the FD,
$E_\text{FD}$ . This ensures that most of the uncertainties associated with
the unknown primary composition in data and hadronic model assumed, as well as
many uncertainties associated with the reconstruction, are absorbed in this
robust and reliable calibration procedure based only on data.

The FD provides a nearly calorimetric, model-independent energy measurement,
because the fluorescence light is produced in proportion to the
electromagnetic energy deposited by the shower in the atmosphere. The
longitudinal profile of the energy deposit, $\text{d}E/\text{d}X$, is determined from the
signals at the triggered pixels of the telescope cameras after taking into account
the separate fluorescence and Cherenkov light contributions, as well as light
attenuation and dispersion in the atmosphere~\cite{FDRecons} . The
electromagnetic energy released by the shower in the atmosphere is obtained by
fitting the longitudinal profile $\text{d}E/\text{d}X$ to a Gaisser-Hillas  (GH)
function~\cite{GH} and integrating over the $X$-range.  The total primary
energy, $E_\text{FD}$, is then derived by adding the so-called ``invisible
energy'' carried into the ground by high-energy muons and neutrinos, which is
estimated using an unbiased and model-independent
method~\cite{TuerosICRC2013}.  The shower-energy estimated with the FD has a
total systematic uncertainty of 14\%~\cite{AbsFDunc}.

For the calibration, only events with zenith angles greater than $60^\circ$
independently triggered and reconstructed by the FD and SD are considered.
The FD measurements must pass quality cuts designed to select high-quality
longitudinal profiles observed under good atmospheric conditions, including
the condition that the depth of the shower maximum, $X_\text{max}$, is within
the field of view (FOV) of the telescopes. The latter condition together with the
limited FOV of the FD may introduce a different selection efficiency for
different primary masses. To avoid this effect, a strict cut on the slant
depth range observed by the telescopes is also applied to ensure that the FOV
is large enough to observe all plausible values of the depth of the shower
maximum for the geometry of each individual shower.  See~\cite{InclinedRec}
for a full description of the selection criteria.  Finally, only events with
energies above $4{\times}10^{18}$\,eV are accepted to ensure practically full
trigger efficiency (see Section~\ref{sec:exposure}).

Here we present the calibration results using inclined events with zenith
angles between $60^\circ$ and $80^\circ$ recorded from 1 January 2004 to 30
September 2013, increasing the data sample by about 16\% with respect to the
one used in previous analyses~\cite{InclinedRec,MuonNumberMeasurement}.  A
total of 255 hybrid events are selected.

\begin{figure}[t]
\centering
\includegraphics[width=0.7\textwidth]{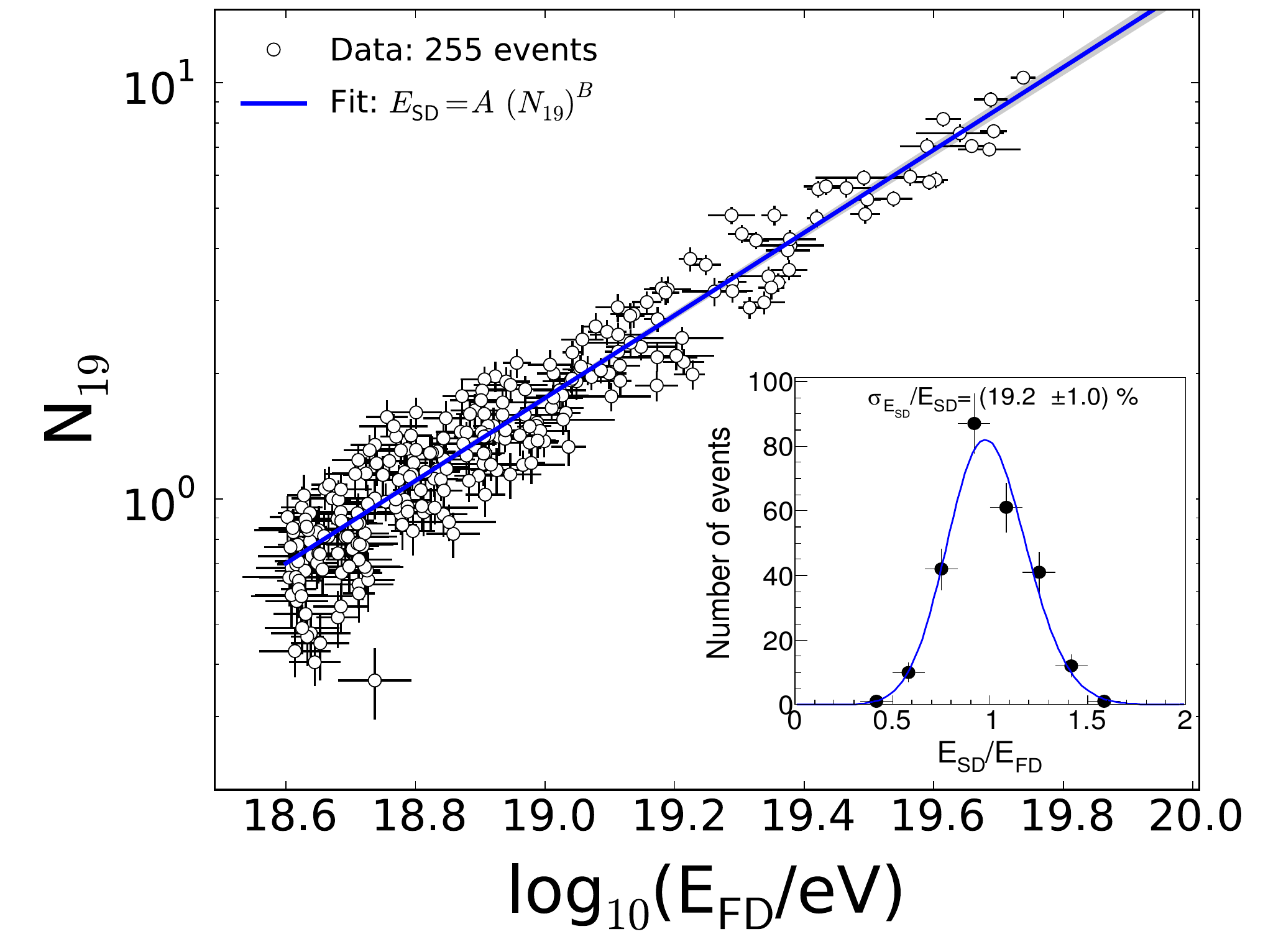}
\caption{Correlation between the shower size parameter, $N_{19}$, and the
  reconstructed FD hybrid energy, $E_\text{FD}$, for the selected hybrid data
  with $\theta\geqslant60^\circ$ used in the fit. The uncertainties indicated
  by the error bars are described in the text. The solid line is the best fit
  of the power-law dependence $E_\text{SD}=A\,(N_{19})^B$ to the data. The
  corresponding ratio distribution of the SD energy $E_\text{SD}$ to the FD
  energy $E_\text{FD}$ is shown in the inset.}
\label{fig:calib}
\end{figure}

The correlation between the energy estimator $N_{19}$ and the calorimetric
hybrid energy $E_\text{FD}$ is well described by a simple power-law function:
\begin{equation}
 N_{19}=A^\prime\,(E_\text{FD}/10^{19}\,\text{eV})^{B^\prime}.
\label{correlation} 
\end{equation}
 The fit is based on a tailored maximum-likelihood method~\cite{HansCalib}
 that includes both the uncertainties of $N_{19}$ and $E_\text{FD}$ without
 relying on approximations.  For each event the uncertainty in $N_{19}$
 corresponds to the reconstruction uncertainty as obtained from the fit to the
 reference muon distribution.  Shower-to-shower fluctuations of $N_{19}$ also
 contribute to the spread and are taken into account. The relative uncertainty
 associated with shower-to-shower fluctuations is assumed to be constant in the
 energy range of interest and fitted to the data as an extra parameter.  The
 uncertainty in $E_\text{FD}$ assigned to each event comes from the
 propagation of both the statistical uncertainties of the reconstruction
 procedure (associated with the reconstruction of the shower geometry, the GH
 fit to the longitudinal profile and the correction for the invisible energy)
 and the uncertainties of the atmospheric parameters~\cite{AbsFDunc}.

The resulting fit to the selected data is illustrated in
figure~\ref{fig:calib}. By inverting the fitted function, the SD energy
estimate is given as $E_\text{SD} = A\,(N_{19})^B$ and the corresponding
calibration parameters are $A=(5.701\pm0.086){\times}10^{18}$\,eV and
$B=1.006\pm0.018$.  Although statistical uncertainties of the calibration
constants $A$ and $B$ affect the SD energy scale, their contribution is small
(from 1.5\% at $10^{19}$\,eV to 4.5\% at $10^{20}$\,eV), decreasing as the
number of events increase.  There is also an uncertainty of ${\sim}2\%$
attributed to the different angular distributions of the hybrid events and the
full inclined data set used to calculate the spectrum. The main contribution
to the systematic uncertainty comes from the overall systematic uncertainty of
14\% from the FD energy scale. By adding the uncertainties described above in
quadrature, the total systematic uncertainty in $E_\text{SD}$ ranges between
14\% at $10^{19}$\,eV and 17\% at $10^{20}$\,eV.

The resolution in $E_\text{SD}$ is computed from the distribution of the ratio
of $A\,(N_{19})^B/E_\text{FD}$ for the hybrid events used for the calibration,
assuming a fixed FD energy resolution of 7.6\%~\cite{AbsFDunc} (method based
on~\cite{Geary}). The resulting average SD energy resolution is estimated to
be $(19\pm1)\%$ (see inset in figure~\ref{fig:calib}). It is dominated by
low-energy showers due to the limited number of events at the highest energies.

\section{Efficiency and exposure}
\label{sec:exposure}

The final step in measuring the energy spectrum is a precise determination of
the exposure for the observations. The exposure involves the time integral of
the aperture, that is, the integral of the instantaneous effective area of the
SD over solid angle and observation time. This integral is subsequently
weighted by the trigger and event selection efficiency, which depends on the
characteristics of the shower such as the nature of the primary particle, its
energy and arrival direction. For low energies, the efficiency is smaller than
one due to the trigger and the selection procedures (details
in~\cite{InclinedRec,Piera}) and can in practice be a source of large
uncertainty. It is thus advantageous to limit the spectral measurements to
energies at which the array is fully efficient, that is, when the effective
area of the SD coincides with the geometrical one.

The inclined SD data set used to measure the cosmic-ray spectrum is composed
of events selected to have zenith angles between $60^\circ$ and $80^\circ$ and
falling inside an active region of the array to guarantee that no crucial part
of the shower is missing. Hence, the trigger and event selection efficiency of
the SD array for inclined showers is the probability that a cosmic-ray shower
reaching an active region of the array induces a trigger, is selected and
finally reconstructed.

\begin{figure}[t]
\centering
\includegraphics[width=0.7\textwidth]{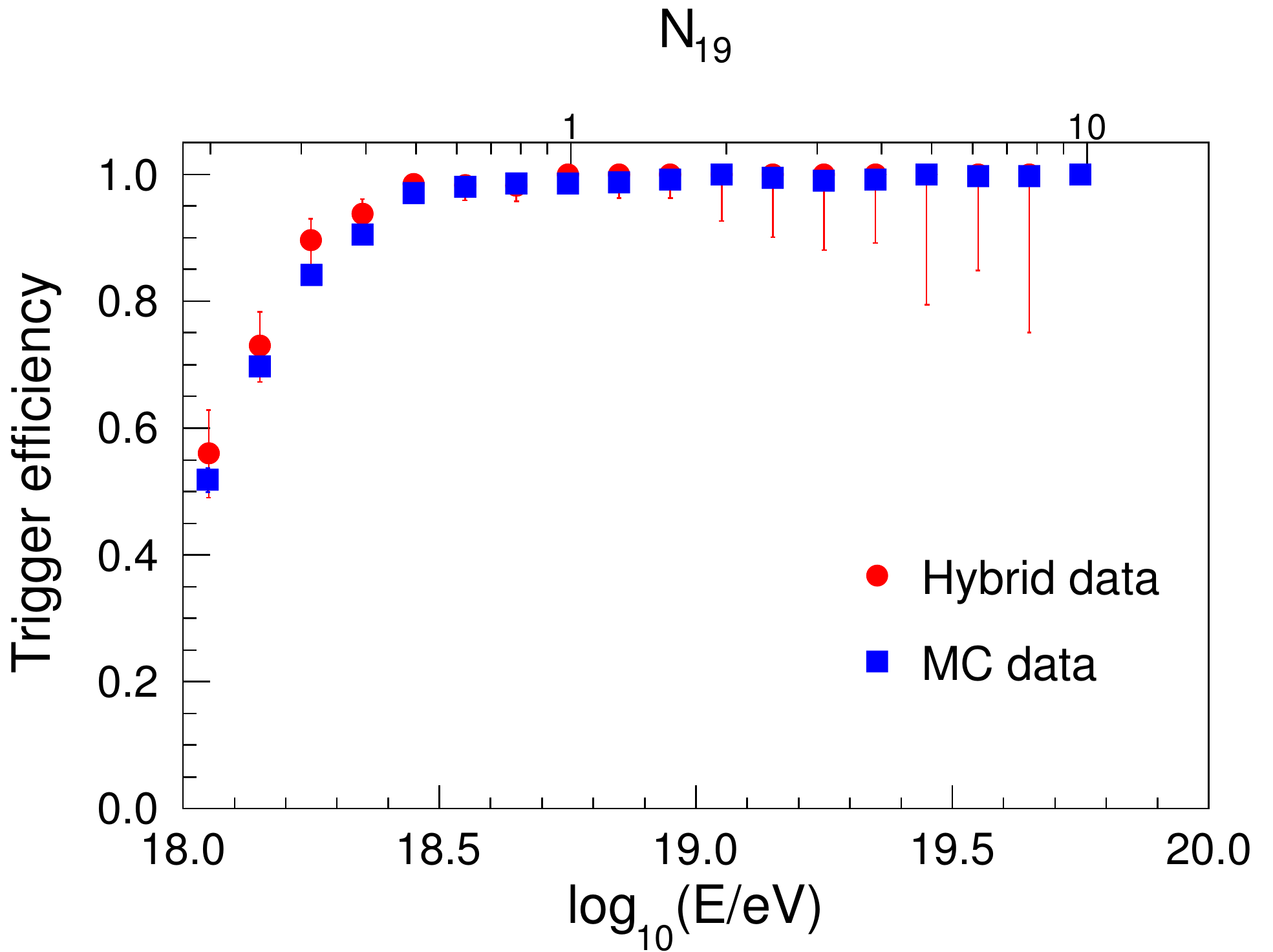}
\caption{Trigger and event selection efficiency of the SD array for showers
  with zenith angles between $60^\circ$ and $80^\circ$ as a function of shower
  energy derived from the hybrid data (circles) and from the Monte Carlo
  simulated showers (squares). The error bars indicate the statistical
  uncertainty (the 68\% probability contour).}
\label{fig:efficiency}
\end{figure}

The energy for full detector efficiency is determined using hybrid events
(detected with the FD and having at least one triggered SD station). To avoid
biases from the primary composition, the same data selection criteria as for
the energy calibration are applied (see
Section~\ref{sec:reconstruction}). Additionally, it is required that the
shower core reconstructed with the FD technique falls within an active cell of
the SD array. Assuming that the detection probabilities of the SD and FD
detectors are independent, the average detection efficiency as a function of
calorimetric shower energy reconstructed with the FD can be estimated from the
fraction of hybrid events that trigger the SD, are selected and finally
reconstructed~\cite{InclinedRec}. The resulting efficiency is shown in
figure~\ref{fig:efficiency} as a function of both shower energy $E$ and shower
size parameter $N_{19}$. The conversion of $N_{19}$ to energy has been made
using the calibration procedure based on data discussed in the
Section~\ref{sec:reconstruction}. The efficiency evaluated with the hybrid
approach is subject to relatively large uncertainties, indicated as error
bars, which are due largely to the limited number of events. The trigger
efficiency of the SD array is found to be almost fully efficient (${>}98\%$)
for energies above $4{\times}10^{18}$\,eV or for $N_{19}>0.7$.  This value
will be used as our estimate of the energy for full efficiency.

The energy for full efficiency obtained with the hybrid apprach is
cross-checked using full shower and detector simulations. The sample used
consists of about 20\,000 proton showers simulated with
\textsc{Corsika}~\cite{Corsika} using \textsc{QGSJetII-04}~\cite{QGSJET-04}
with zenith angles isotropically distributed between $60^\circ$ and $80^\circ$
and energies ranging from $\log_{10}(E/\text{eV})=18$ to 20, in steps of 0.5
(with a spectral index $\gamma=1$ in each sub-interval). To generate the
simulated events, these showers subsequently underwent a full simulation of
the detector response, within the
\mbox{$\overline{\textrm{Off}}$\hspace{.05em}\protect\raisebox{.4ex}{$\protect\underline{\textrm{line}}$}}
framework, with random impact points in the SD array. Only showers with impact
points in an active region of the array are considered for this analysis. For
each event, the full trigger and event selection chain are applied.

The trigger efficiency for inclined showers depends mainly on the total number
of muons at the ground for a given arrival direction, which is proportional to
the reconstructed shower size parameter $N_{19}$.  A recent
study~\cite{MuonNumberMeasurement} has reported that the measured muon number
in inclined showers at ground level exceeds expectations obtained with
simulations and various hadronic interaction models (even when assuming a pure
iron composition).  This implies that simulations and data have a different
$N_{19}{-}E$ correlation, that is, the MC energy is not directly comparable to
$E_\text{FD}$ for the same muon content.  To carry out a comparison between
using hybrid data and simulations, we correct for the muon deficit observed in
simulations by converting the shower size of each simulated event into energy
using the same calibration procedure applied for the data.  The resulting
efficiency as a function of both shower energy and shower size parameter is
also shown in figure~\ref{fig:efficiency}.  The trigger and selection become almost
fully efficient at energies above $4{\times}10^{18}$\,eV in agreement with
hybrid data. It has been tested with simulations that above this
energy, the detector is fully efficient independently of the particular choice
of mass composition and hadronic model.

The choice of a fiducial trigger based on active hexagons allows us to exploit
the regularity of the array~\cite{Piera}. The geometrical aperture of the
array is obtained as a multiple of the aperture of an elemental hexagon
cell. Above the energy for full efficiency, the detection area per cell is
1.95\,km$^2$, which results in an aperture of $a_\text{cell}=1.35$\,km$^2$\,sr
for showers with zenith angles between $60^\circ$ and $80^\circ$. The
calculation of the integrated exposure over a given period of time simply
amounts to counting the operational cells as a function of time,
$N_\text{cell}(t)$, and integrating the corresponding aperture,
$N_\text{cell}(t) \times a_\text{cell}$, over time.  The overall uncertainty
on the integrated exposure is less than $3\%$~\cite{Piera}.

In January 2004 the initial area spanned by the array was about 34\,km$^2$ and
it rose steadily until August 2008 when over 1600 surface detectors were in
operation.  The calculation of the surface area is the same as that used for
the spectrum measured with events at zenith angles less than $60^\circ$ and
has been done numerically, making use of the trigger rate at each
station~\cite{Piera}. The integrated exposure for showers with zenith angles
between $60^\circ$ and $80^\circ$ amounts to ($10\,890 \pm
330$)\,km$^2$\,sr\,yr in the time period from 1 January 2004 to 31 December
2013 (which corresponds to 29\% of the exposure for showers with
$\theta{<}60^{\circ}$ in the same period). This calculation excludes periods
during which the array was not sufficiently stable, which add up to a small
fraction of the total time (of the order of 5\%).

\section{Spectrum}
\label{sec:spectrum}

Inclined events recorded by the SD in the time period between 1 January 2004
and 31 December 2013 were analysed and calibrated with the procedure outlined
above.  The resulting data set consists of 254\,686 events that fell on the
active part of the array with zenith angles in the interval between $60^\circ$
and $80^\circ$.  To avoid large uncertainties due to the trigger and event
selection efficiencies, only events having energies greater than
$4{\times}10^{18}$\,eV are considered for the spectrum calculation, reducing
the sample to 15\,614 events.

\begin{figure}[t]
\centering
\includegraphics[width=0.8\textwidth]{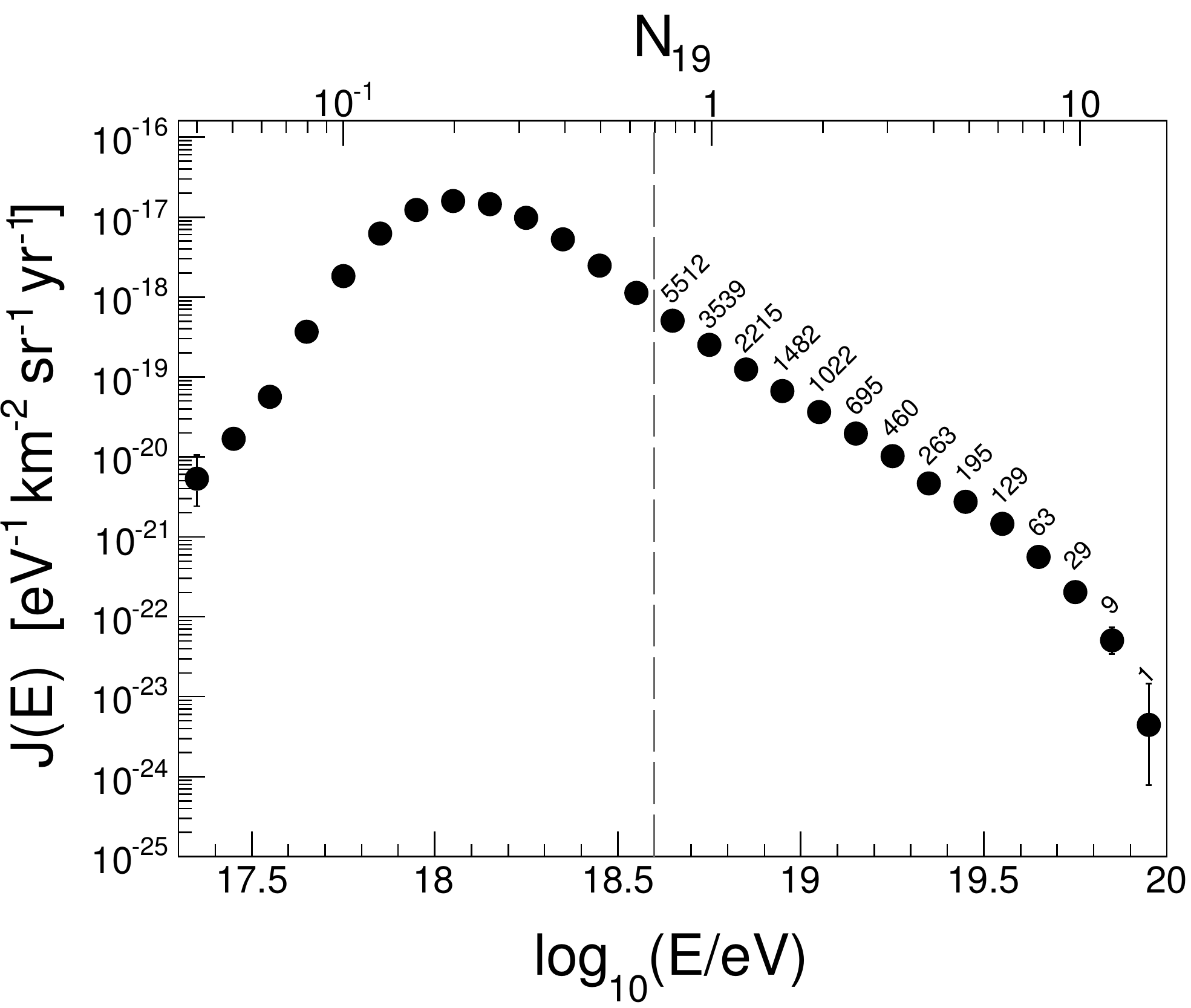}
\caption{Uncorrected energy spectrum of the cosmic rays derived from inclined
  events with zenith angles $60^\circ\leqslant\theta\leqslant80^\circ$ in
  terms of energy and shower size. Only statistical uncertainties are
  shown. The number of events in each of the bins above $4{\times}10^{18}$\,eV
  (vertical dashed line) is given above the points.}
\label{fig:RawSpectrum}
\end{figure}

The differential flux of cosmic rays at a certain energy $E$, $J(E)$,
is obtained by dividing the energy distribution of the cosmic rays by
the accumulated exposure in the corresponding zenith-angle interval
for the energies in the range of full trigger
efficiency. Figure~\ref{fig:RawSpectrum} shows this ratio for the
selected events (also displayed as a function of the equivalent shower
size $N_{19}$). This is a raw distribution since effects of the finite
resolution of the SD energy measurement have not yet been taken into
account. Note that this figure also shows the ratio below
$4{\times}10^{18}$\,eV where it is not expected that the flux will be
reproduced accurately because of the energy dependence of the
triggering efficiency (see figure~\ref{fig:efficiency}).

%
%
%

A correction must be applied to account for the effect of the resolution in
the energy determination, responsible for a bin-to-bin event migration. The
number of events in a given bin is contaminated by movements of reconstructed
energies from the adjacent bins. For an energy spectrum which is steeply
falling, upward fluctuations into a given bin are not completely compensated
by other fluctuations from the other direction, and the net effect is an
overestimate of the flux when this correction is not applied. Due to this, the
measured spectrum is shifted towards higher energies with respect to the true
spectrum.  To correct for this, a forward-folding approach is applied. Monte
Carlo simulations are used to determine the resolution of the SD energy
estimator based on the shower size parameter, $N_{19}$, for different
assumptions on the primary mass and hadronic interaction models. Then the
average resolution is converted to the SD energy resolution using the same
energy scale as obtained from the data. In addition, intrinsic fluctuations
(also called shower-to-shower fluctuations) contribute to the bin
migration. In this case, these fluctuations are modelled with a normal
distribution that has a constant relative standard deviation,
$\sigma(N_{19})/N_{19}$, and are estimated in the data-calibration procedure
as explained in~\cite{InclinedRec}, resulting in $\sigma(N_{19})/N_{19} =
(14\pm1)\%$. From the SD energy resolution and intrinsic fluctuations, a
bin-to-bin migration matrix is derived. The matrix is then used to find a flux
parameterisation that matches the measured data when forward-folded.

%
%
\begin{figure}[t]
\centering
\includegraphics[width=0.65\textwidth]{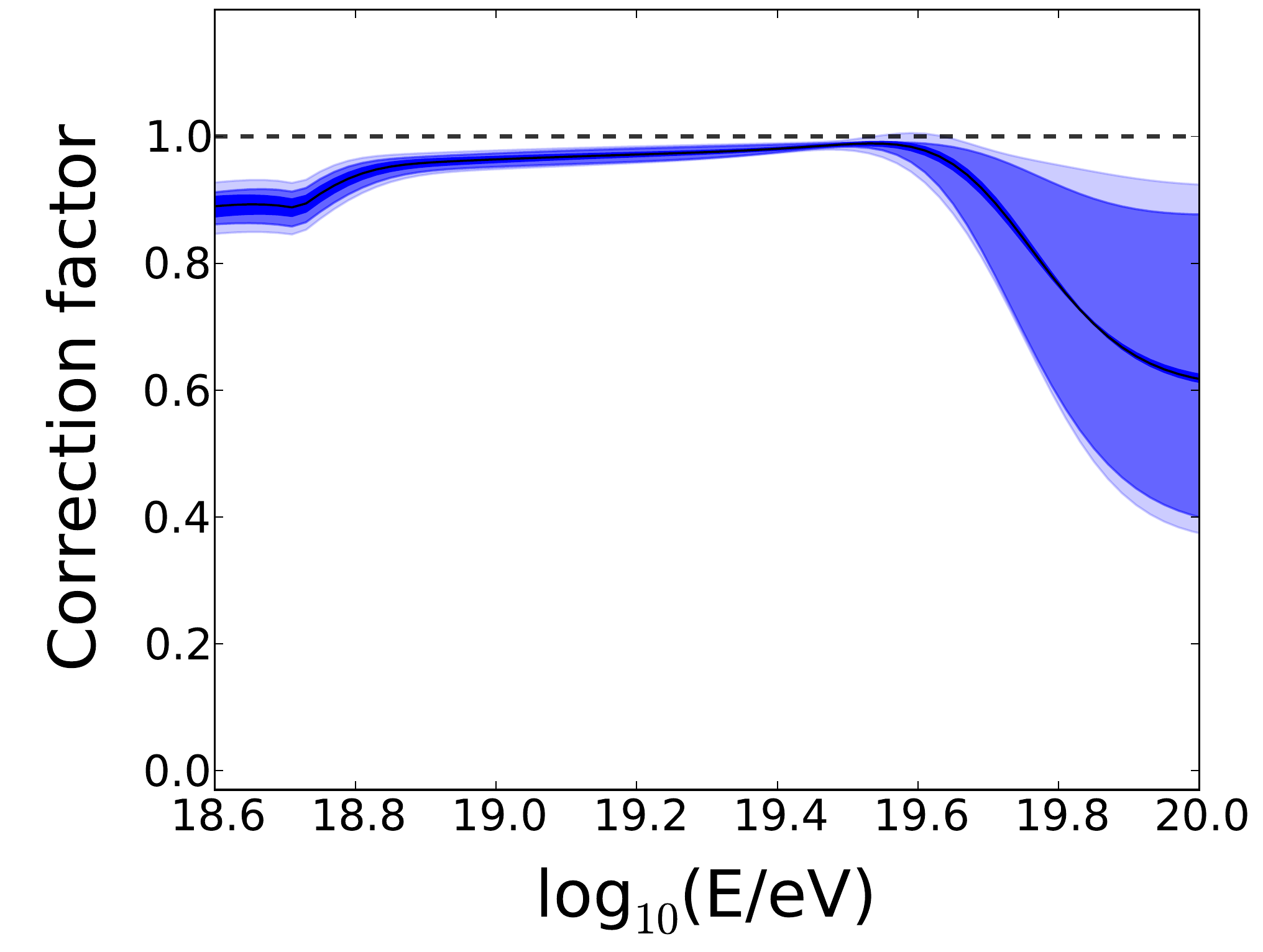}
\caption{Correction factor applied to the measured spectrum to account for the
  detector effects as a function of the cosmic-ray energy. The uncertainty due
  to the energy resolution is shown with the dark band, and that
  due to shower-to-shower fluctuations with the blue band. The
  total uncertainty, also including the uncertainty of the fit, is shown by
  the light band. }
\label{fig:CorrectionFactor}
\end{figure}

We assume a common parameterisation for the spectrum given a power law below
the ankle and a power law plus a smoothly changing function above,
\begin{alignat}{2}
J(E) &\propto E^{-\gamma_1} & \quad ;\,E<E_\text{ankle},
\label{eq:FormBeforeAnkle}
\\
J(E) &\propto E^{-\gamma_2} \, 
  \left[1 + \left(\frac{E}{E_\text{s}}\right)^{\Delta\gamma}\right]^{-1}
  & \quad ;\,E\geqslant E_\text{ankle},
\label{eq:FormAfterAnkle}
\end{alignat}
%
%
%
where $E_\text{s}$ is the energy at which the flux falls to one-half
of the value of the power-law extrapolation of the intermediate region
and $\Delta\gamma$ gives the increment of the spectral index beyond the
suppression region\footnote{Note that $\Delta\gamma=1/\ln(W_\text{c})$
  in the equivalent expression used in~\cite{HybridSpectrum}, where
  $W_\text{c}$ defines the width of the transition region at the suppression.}.  This
parameterisation was convolved with the migration matrix and the
resulting flux was fitted to the measured spectrum, using a binned
maximum-likelihood approach assuming Poisson statistics. As
$E_\text{ankle}$ is very close to the saturation energy for inclined
showers, neither $E_\text{ankle}$ nor $\gamma_1$ can be reliably
established from the data. These parameters are relevant for the
unfolding of the lower-energy part of the spectrum. They were fixed to
the values obtained in the spectrum reported in \cite{SchulzICRC2013},
$\gamma_1=3.23\pm0.01\pm0.07\,\text{(sys)}$ and $E_\text{ankle} =
(5.25\pm0.12\pm0.24\,\text{(sys)}){\times}10^{18}$\,eV , while the
other parameters were left free.  The convolved flux is finally
divided by the input flux to obtain the correction factors which are
in turn applied to the measured binned spectrum.

The correction factor resulting from the fit of the assumed parameterisation
is shown in figure~\ref{fig:CorrectionFactor}. The uncertainty from the fit
was obtained by propagating the covariance matrix of the fitted parameters
into the correction function.  A systematic uncertainty is attributed to
possible variations of the shower-to-shower fluctuations, which can arise if
the mass component changes and because of the changing $X_\text{max}$ over the
energy range of interest.  Although recent
results~\cite{MuonNumberMeasurement} favour a transition from lighter to
heavier elements in the energy range considered, here we assume a conservative
scenario where the relative fluctuations are allowed to vary between 4\%
(corresponding to a pure iron composition) and 21\% (corresponding to a pure
proton composition) over the full energy range. Finally, the propagated
uncertainty in the average resolution of the energy estimator $N_{19}$,
estimated to be on the order of 10\%, was also included.  The uncertainty in
the correction factor that arises from the uncertainties in the ankle
parameters ($E_\text{ankle}$ and $\gamma_1$), fixed to the values reported
in \cite{SchulzICRC2013}, is negligible ($<$1\% below $E_\text{s}$).  In
figure~\ref{fig:CorrectionFactor}, in addition to the total uncertainty (light
band), the last two uncertainty components are shown separately to illustrate
that the main contribution is the systematic uncertainty due to variations of
the shower-to-shower fluctuations.

%
\begin{figure}[t]
\centering
\includegraphics[width=0.8\textwidth]{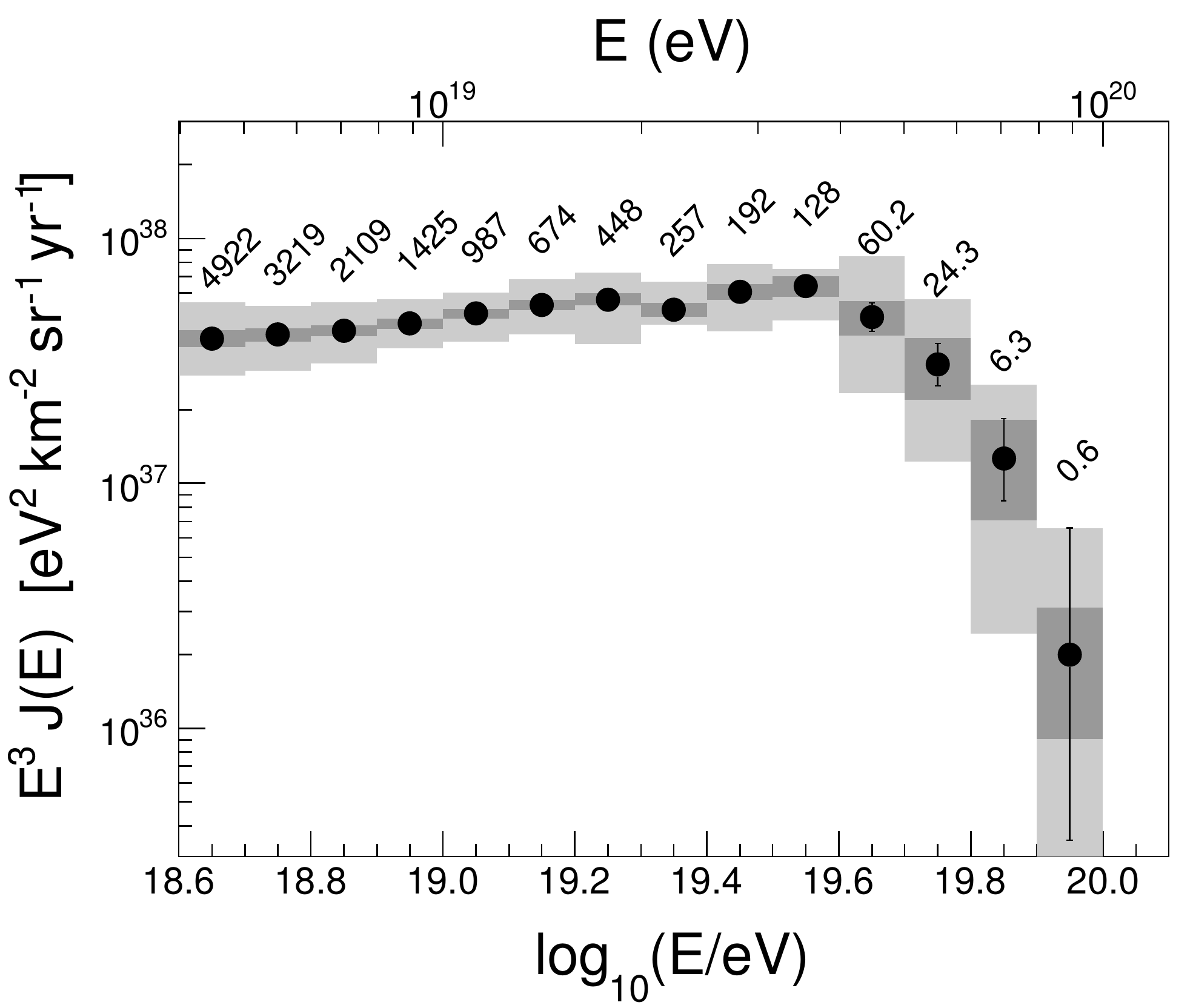}
\caption{Energy spectrum of the cosmic rays, corrected for energy resolution,
  derived from inclined data with zenith angles
  $60^\circ\leqslant\theta\leqslant80^\circ$. The error bars represent
  statistical uncertainties. The light shaded boxes indicate the total
  systematic uncertainties. For illustration purposes, the systematic
  uncertainties excluding the uncertainty from the energy scale are also shown
  as darker boxes. The effective number of events, after correcting the flux
  for energy resolution, is given above the points.}
\label{fig:Spectrum}
\end{figure}
%

%
\begin{figure}[t]
\centering
\includegraphics[width=0.7\textwidth]{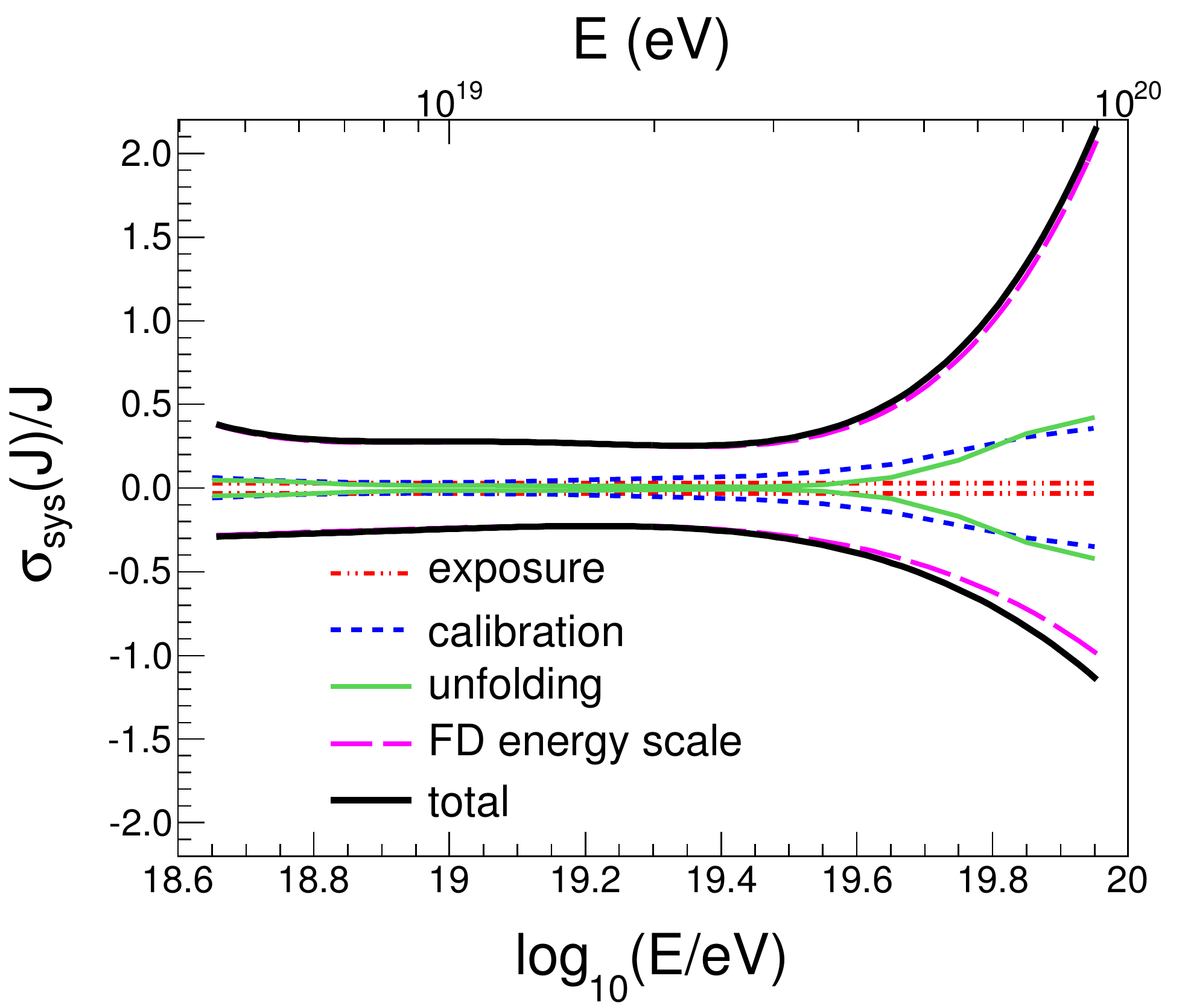}
\caption{Systematic uncertainties in the derived flux as a function of
  the cosmic-ray energy. For a description of the different sources of
  systematic uncertainty, see the text.}
\label{fig:SysUnc}
\end{figure}
%

The spectrum, corrected for energy resolution, is shown in
figure~\ref{fig:Spectrum}. The error bars represent statistical uncertainties
only. The flux is multiplied by $E^{3}$ to better present its features, a flat
region above $4{\times}10^{18}$\,eV up to the steepening at energies above
about $4{\times}10^{19}$\,eV. The light shaded boxes indicate the total
systematic uncertainties (less than ${\sim}40\%$ up to $4{\times}10^{19}$\,eV,
and then increasing up to ${\sim}200\%$ for the highest energy bin), which
include the uncertainty in the calibration parameters propagated to the flux,
a global uncertainty derived from the SD exposure calculation (3\%), the
uncertainty arising from the unfolding process, and the global systematic
uncertainty of the FD energy scale (14\%) from the hybrid-calibration
procedure. In figure~\ref{fig:SysUnc} these separate contributions to the
systematic uncertainties in the derived flux are shown as a function of the
cosmic-ray energy.

\section{Discussion}
\label{sec:discussion}

\begin{figure}[t]
\centering
\includegraphics[width=0.8\textwidth]{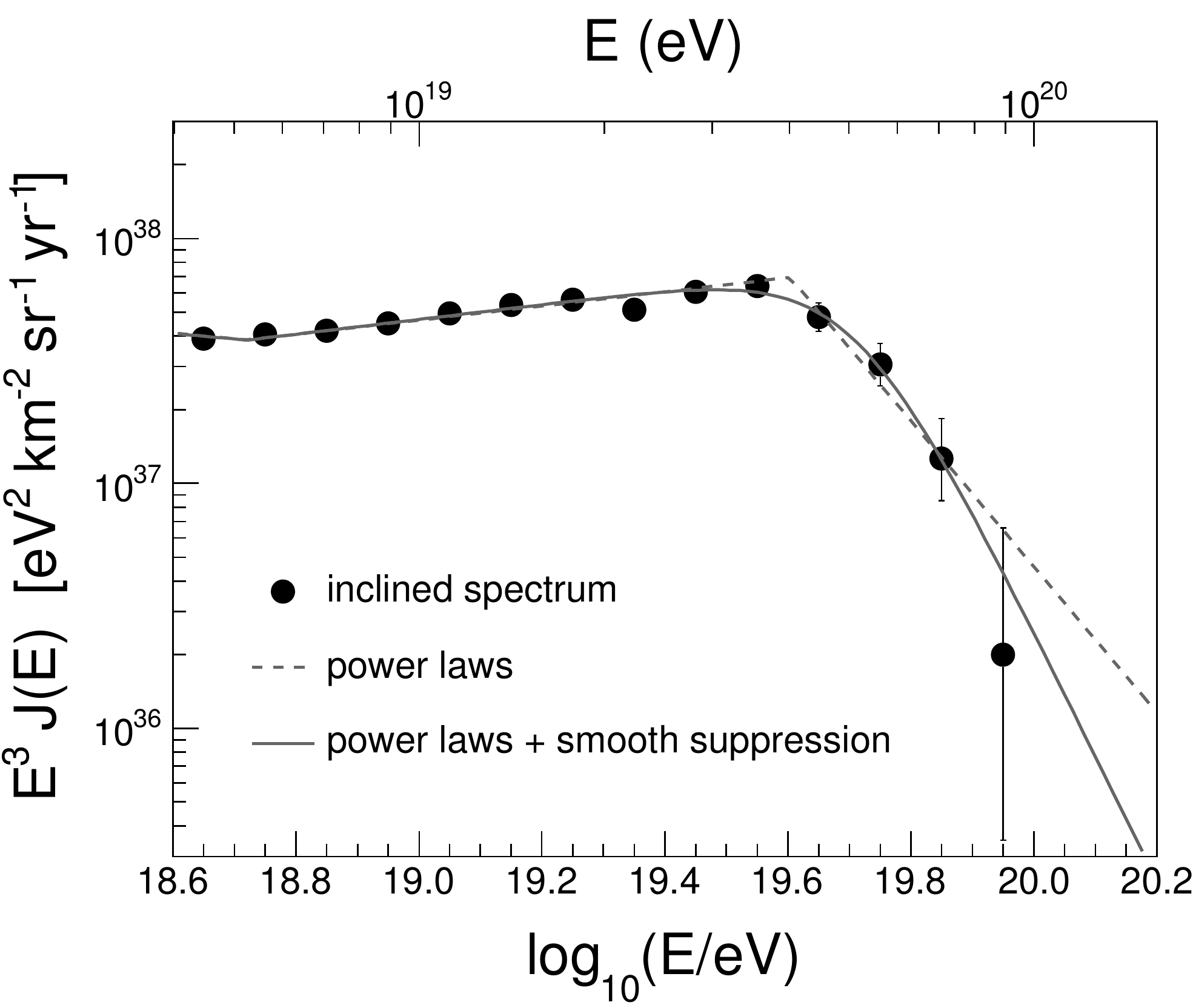}
\caption{Energy spectrum of the cosmic rays, corrected for energy resolution,
  derived from inclined data fitted with simple power laws (dashed
  line) and with eqs.~\eqref{eq:FormBeforeAnkle} and~\eqref{eq:FormAfterAnkle}
  (solid line). The systematic uncertainty on the energy scale is 14\%.}
\label{fig:SpectrumFit}
\end{figure}

The characteristic features of the spectrum have been quantified by
fitting the model given by eqs.~\eqref{eq:FormBeforeAnkle}
and~\eqref{eq:FormAfterAnkle}, which assumes a spectrum with a sharp
ankle and a smooth suppression at the highest energies, to the
unfolded spectrum. The result of the best fit is shown as a solid line
in figure~\ref{fig:SpectrumFit}. Another approach is to describe the
spectrum with three power laws separated by two breaking points as
illustrated by a dashed line in figure~\ref{fig:SpectrumFit}. The
first model improves marginally the description of the abrupt change
of slope observed at higher energies. The spectral parameters from the
best fits to the data are given in table~\ref{tb:FittedParameters},
quoting both statistical and systematic uncertainties.


\begin{table*}[!ht]
\centering
\begin{tabular}{lll}
\toprule
Parameter & Power laws & \; Power laws + smooth suppression
\\
\midrule
$\gamma_2(E>E_\text{ankle})$ & $2.71\pm0.02\pm0.1$\,(sys) & \; $2.70\pm0.02\pm0.1$\,(sys)
\\
$E_\text{break}$ & $(4.01\pm0.21^{+1.0}_{-1.7}\,\text{(sys)}){\times}10^{19}$\,eV &
\\
$\gamma_3(E>E_\text{break})$ & $5.98\pm0.61\,^{+0.9}_{-1.4}$\,(sys) &
\\ 
$E_\text{s}$ & & \;$(5.12\pm0.25^{+1.0}_{-1.2}\,\text{(sys)}){\times}10^{19}$\,eV
\\ 
$\Delta\gamma$ & &\; $5.4\pm1.0^{+2.1}_{-1.1}$\,(sys)
\\
$\chi^2$/ndof & 15.7/10 & \; 13.2/10
\\
\bottomrule
\end{tabular}
\caption{Fitted parameters, with statistical and systematic uncertainties,
  parameterising the energy spectrum measured with the inclined events.}
\label{tb:FittedParameters}
\end{table*}

Note that, as explained above, the selected energy threshold of
$4{\times}10^{18}$\,eV is, by coincidence, close to the value obtained for the
ankle, $(5.25\pm0.12){\times}10^{18}$\,eV, in the combined analysis of hybrid
events and SD events~\cite{SchulzICRC2013}. It is thus not possible to analyse
the ankle transition properly. Nevertheless, the raw data do clearly display
such a feature (see figure~\ref{fig:RawSpectrum}) which we do not analyse any
further here due to the large uncertainty in the efficiency in this region
(see figure~\ref{fig:efficiency}). As a consequence, the $\gamma_1$ and
$E_\text{ankle}$ parameters in equation~\eqref{eq:FormBeforeAnkle} have been
fixed to the values reported in \cite{SchulzICRC2013}.

Both models consistently describe the ``flat'' region of the spectrum above
the ankle up to the observed onset of the suppression at
${\sim}4{\times}10^{19}$\,eV by a power law with a spectral index of
$\gamma_{2}=2.7$.  The spectral index after the steepening is less certain due
to the low number of events and large systematic uncertainties. The
significance of the suppression\footnote{Following the recipe given
  in~\cite{Hague:2006is}, the null hypothesis that the power law with spectral
  index $\gamma_2=2.71\pm0.02$ continues beyond the suppression point can be
  rejected due to the low probability $(5\,^{+3}_{-2}){\times}10^{-11}$.} is
${\sim}6.6\,\sigma$ with 220.4 events expected above the break energy
$E_\text{break}$ while only 102 events were observed.

%

A different observable that characterises the suppression is the energy
$E_\text{1/2}$ at which the integral spectrum drops by a factor of two
relative to the power-law extrapolation from lower energies, as suggested
in~\cite{Berezinsky:2002nc}. Here, the integral spectrum was computed by
integrating the parameterisation given by eqs.~\eqref{eq:FormBeforeAnkle}
and~\eqref{eq:FormAfterAnkle} and the parameters reported in
table~\ref{tb:FittedParameters}. The result yields $E_\text{1/2} =
(3.21\pm0.01\pm0.8\,\text{(sys)}){\times}10^{19}$\,eV, which is smaller than
the value of $\approx 5.3 {\times}10^{19}$\,eV predicted for the GZK energy
cutoff of protons~\cite{Berezinsky:2002nc} (practically independent of the
generation index of the assumed energy spectrum), with a difference at the
level of $2.6\,\sigma$. In~\cite{Berezinsky:2002nc} the assumption was made
that the sources of ultra-high energy cosmic rays are uniformly distributed
over the universe and are accelerators of protons. In reality, sources are
discrete and in the GZK region the shape of the spectrum will be dominated by
the actual distribution of sources around us. Further discussion of this point
is given in~\cite{Ahlers:2012az}.  Other scenarios for the high-energy
suppression of the spectrum (e.g.,
\cite{Aloisio:2009sj,Allard:2011aa,Biermann:2011wf,Taylor:2013gga}) attribute
it to the limiting acceleration energy at the sources rather than to the GZK
effect, providing a good description of the combined Auger energy spectrum as
shown in~\cite{Kampert:2014eja}.

%
\begin{figure}[!ht]
\centering
\includegraphics[width=0.8\textwidth]{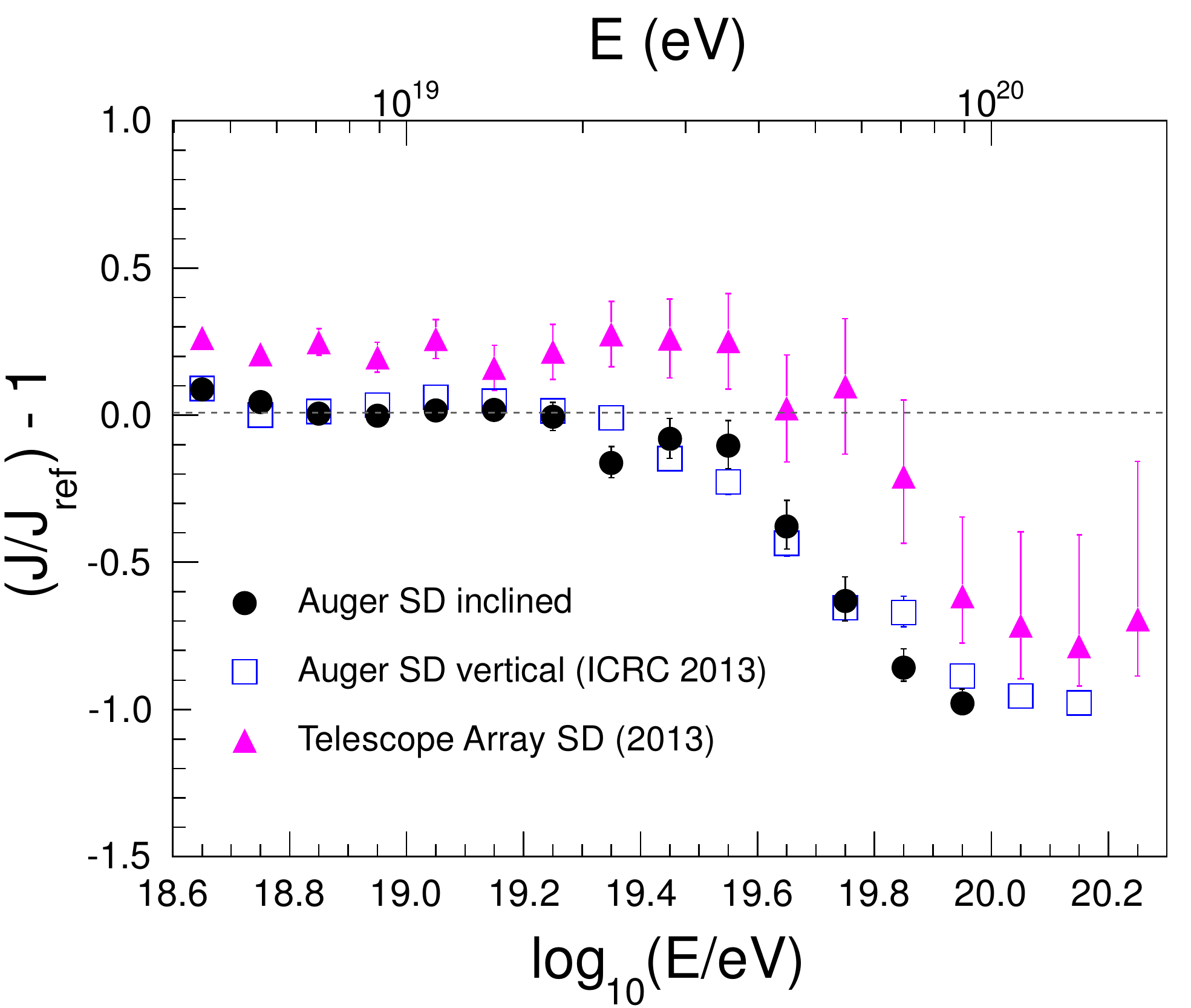}
\caption{Fractional difference between the energy spectrum of cosmic rays
  derived from SD data with $\theta\geqslant60^\circ$ recorded at the Pierre
  Auger Observatory and a reference spectrum with an index of 2.67.
  Residuals for spectra derived from Auger SD data with
  $\theta < 60^\circ$~\cite{SchulzICRC2013} and the Telescope Array SD
  data with $\theta<55^\circ$~\cite{TAspectrum} are also shown
  for comparison.}
\label{fig:ResidualSpectra}
\end{figure}

To compare the spectrum obtained here with other measurements, we have adopted
the technique suggested in~\cite{Watson:2013cla} in which the differential
flux at each energy is compared with the expected differential flux from a
reference spectrum. We choose as reference a spectrum with an
index~\footnote{The index of the reference spectrum has no physical
  significance, so its choice is quite arbitrary.} of 2.67 fitted to the flux
presented here (see figure~\ref{fig:Spectrum}) in the energy bin corresponding
to $\log_{10}(E/\text{eV})=18.95$ (bin width of 0.1), which contains over 1425
events.  The reference spectrum is thus
\begin{equation}
\ J_\text{exp} = 2.52{\times}10^{31}\,E^{-2.67} \,\text{eV}^{-1}\,\text{km}^{-2}\,\text{sr}^{-1}\,\text{yr}^{-1}.
\label{refspectrum}
\end{equation}

In figure~\ref{fig:ResidualSpectra} the spectrum obtained with
inclined events is displayed as a fractional difference (also called the
residual) with respect to the reference spectrum, in comparison to the
residual for the spectrum obtained from data with zenith angles less
than $60^\circ$~\cite{SchulzICRC2013}. The comparison shows that both
spectra are in agreement within errors. We note that the last two
points are systematically below the corresponding measurement
obtained for Auger events with zenith angles less than $60^\circ$. Also
the point corresponding to $\log_{10}(E/\text{eV})=19.35$ is below the
vertical one. However, the differences are at the $2\,\sigma$ level.

%
\begin{figure}[!ht]
\centering
\includegraphics[width=0.8\textwidth]{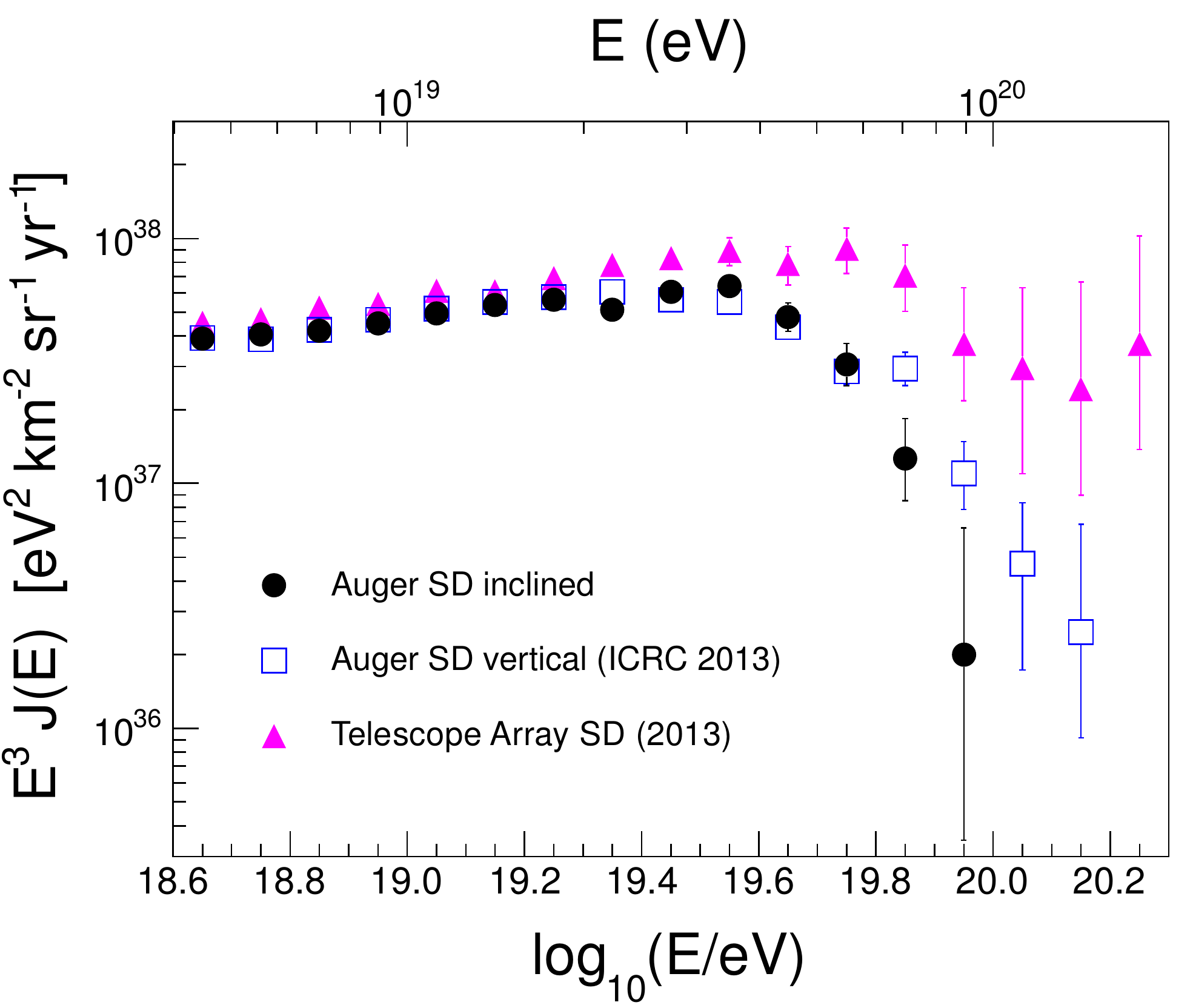}
\caption{Compilation of the energy spectrum of cosmic rays derived from SD
  data recorded at the Pierre Auger Observatory with   $\theta\geqslant60^\circ$ (circles) and $\theta <
  60^\circ$ (squares), and at the Telescope Array with $\theta<55^\circ$ (triangles).}
\label{fig:AllSpectra}
\end{figure}

Figure~\ref{fig:ResidualSpectra} also displays the spectrum obtained from SD
events with zenith angles less than $55^\circ$ recorded with the Telescope
Array (TA) detector~\cite{TAspectrum}. The comparison of the residuals for the
three spectra (also illustrated in the form of $J\,E^{3}$ in
figure~\ref{fig:AllSpectra}) shows that in the region between the ankle and
the suppression the Auger spectra fit well due to our choice of reference
spectrum, and the average of the residuals for TA is about +23\%. The spectra
determined by the Auger and TA Observatories are consistent in normalisation
and shape within the systematic uncertainties in the energy scale quoted by
the two collaborations~\cite{IoanaUHECR2014}. However, differences are clearly
seen in the high-energy region and are not understood thus far. Understanding
the origin of this difference, whether from anisotropies at high energies,
composition-related effects, experimental issues or any other reason, is of
high priority in the efforts to understand the origin of the UHECRs. However
the dedicated study of this discrepancy is beyond the scope of this
work. Since 2012 there has been a collaborative effort between the Pierre
Auger and Telescope Array Collaborations to investigate the level of agreement
between the different measurements of the UHECR energy spectrum and to
understand the sources of possible discrepancies, by examining the different
measurement techniques and analysis methods employed by these groups. The
latest results obtained by the energy spectrum working group can be found
in~\cite{IoanaUHECR2014}.

\section{Summary}

The cosmic-ray spectrum has been obtained for energies exceeding
$4{\times}10^{18}$\,eV using showers with zenith angles between $60^\circ$ and
$80^\circ$ recorded with the surface detector of the Pierre Auger Observatory
during the time period between 1 January 2004 and 31 December 2013. It has
been shown that the SD array becomes fully efficient for inclined events above
this energy. The results can be described by a power-law spectrum with
spectral index 2.7 above $5{\times}10^{18}$\,eV and clearly indicate a
steepening of the cosmic-ray spectrum above an energy around
$4{\times}10^{19}$\,eV. These features and the normalisation of the spectrum
are in agreement with previous measurements made with the Pierre Auger
Observatory (using SD data and hybrid data with zenith angles less than
$60^\circ$) and also with the measurements of the Telescope Array.  

The inclined data set is independent and complementary to the vertical
data set and its reconstruction is performed using an independent
method. These data provide a $29\%$ increase in number of events over
the previous event set. In addition to obtaining an independent
measurement of the cosmic-ray spectrum reported here, the inclined
data are being analysed to explore primary composition, to constrain
the current models that attempt to describe the hadronic interactions
at energies above $4{\times}10^{18}$\,eV
($E_\text{CM}\approx87$\,TeV), and to improve the studies of the
arrival directions of the cosmic rays by extending the accessible
fraction of the sky to 85\%.


\section*{Acknowledgements}

The successful installation, commissioning, and operation of the Pierre Auger
Observatory would not have been possible without the strong commitment and
effort from the technical and administrative staff in Malarg\"{u}e.  We are
very grateful to the following agencies and organizations for financial
support:

 Comisi\'{o}n Nacional de Energ\'{\i}a At\'{o}mica, 
 Fundaci\'{o}n Antorchas, Gobierno de la Provincia 
 de Mendoza, Municipalidad de Malarg\"{u}e, 
 NDM Holdings and Valle Las Le\~{n}as, in gratitude 
 for their continuing cooperation over land access, 
 Argentina; the Australian Research Council; Conselho 
 Nacional de Desenvolvimento Cient\'{\i}fico e 
 Tecnol\'{o}gico (CNPq), Financiadora de Estudos e 
 Projetos (FINEP), Funda\c{c}\~{a}o de Amparo \`{a} 
 Pesquisa do Estado de Rio de Janeiro (FAPERJ), 
 S\~{a}o Paulo Research Foundation (FAPESP) 
 Grants No. 2010/07359-6 and No. 1999/05404-3, 
 Minist\'{e}rio de Ci\^{e}ncia e Tecnologia (MCT), 
 Brazil; Grant No. MSMT-CR LG13007, No. 7AMB14AR005, 
 and the Czech Science Foundation Grant No. 14-17501S, 
 Czech Republic;  
 Centre de Calcul IN2P3/CNRS, Centre National de la 
 Recherche Scientifique (CNRS), Conseil R\'{e}gional 
 Ile-de-France, D\'{e}partement Physique Nucl\'{e}aire 
 et Corpusculaire (PNC-IN2P3/CNRS), D\'{e}partement 
 Sciences de l'Univers (SDU-INSU/CNRS), Institut 
 Lagrange de Paris (ILP) Grant No. LABEX ANR-10-LABX-63, 
 within the Investissements d'Avenir Programme  
 Grant No. ANR-11-IDEX-0004-02, France; 
 Bundesministerium f\"{u}r Bildung und Forschung (BMBF), 
 Deutsche Forschungsgemeinschaft (DFG), 
 Finanzministerium Baden-W\"{u}rttemberg, 
 Helmholtz Alliance for Astroparticle Physics (HAP), 
 Helmholtz-Gemeinschaft Deutscher Forschungszentren (HGF), 
 Ministerium f\"{u}r Wissenschaft und Forschung, Nordrhein Westfalen, 
 Ministerium f\"{u}r Wissenschaft, Forschung und Kunst, Baden-W\"{u}rttemberg, Germany; 
 Istituto Nazionale di Fisica Nucleare (INFN), Ministero dell'Istruzione, dell'Universit\'{a} 
 e della Ricerca (MIUR), Gran Sasso Center for Astroparticle Physics (CFA), CETEMPS Center 
 of Excellence, Ministero degli Affari Esteri (MAE), Italy; 
 Consejo Nacional de Ciencia y Tecnolog\'{\i}a (CONACYT), Mexico; 
 Ministerie van Onderwijs, Cultuur en Wetenschap, 
 Nederlandse Organisatie voor Wetenschappelijk Onderzoek (NWO), 
 Stichting voor Fundamenteel Onderzoek der Materie (FOM), Netherlands; 
 National Centre for Research and Development, 
 Grants No. ERA-NET-ASPERA/01/11 and 
 No. ERA-NET-ASPERA/02/11, National Science Centre,
 Grants No. 2013/08/M/ST9/00322, No. 2013/08/M/ST9/00728 
 and No. HARMONIA 5 - 2013/10/M/ST9/00062, Poland; 
 Portuguese national funds and FEDER funds within 
 Programa Operacional Factores de Competitividade 
 through Funda\c{c}\~{a}o para a Ci\^{e}ncia e a  Tecnologia (COMPETE), Portugal; 
 Romanian Authority for Scientific Research ANCS, 
 CNDI-UEFISCDI partnership projects Grants No. 20/2012 
 and No. 194/2012, Grants No. 1/ASPERA2/2012 ERA-NET, 
 No. PN-II-RU-PD-2011-3-0145-17 and No. PN-II-RU-PD-2011-3-0062, 
 the Minister of National  Education, Programme  
 Space Technology and Advanced Research (STAR), 
 Grant No. 83/2013, Romania; 
 Slovenian Research Agency, Slovenia; 
 Comunidad de Madrid, FEDER funds, Ministerio de Educaci\'{o}n y Ciencia, 
 Xunta de Galicia, European Community 7th Framework Program, 
 Grant No. FP7-PEOPLE-2012-IEF-328826, Spain; 
 Science and Technology Facilities Council, United Kingdom; 
 Department of Energy, 
 Contracts No. DE-AC02-07CH11359, No. DE-FR02-04ER41300, 
 No. DE-FG02-99ER41107 and No. DE-SC0011689, 
 National Science Foundation, Grant No. 0450696, 
 The Grainger Foundation, USA; 
 NAFOSTED, Vietnam; 
 Marie Curie-IRSES/EPLANET, European Particle Physics 
 Latin American Network, European Union 7th Framework 
 Program, Grant No. PIRSES-2009-GA-246806; and UNESCO.

\end{document}